\documentclass{aa}

\usepackage{lscape}
\usepackage{graphicx}
\usepackage{rotating}
\usepackage{natbib}
\usepackage{float}
\newcommand{\barray}{\begin{eqnarray}}
\newcommand{\earray}{\end{eqnarray}}

\begin{document}

\title{Ejection of gaseous clumps from  gravitationally unstable protostellar disks}
\titlerunning{Ejection of gaseous clumps}

\author
{E. I. Vorobyov,$^{1,2}$}
\institute{Department of Astrophysics, University of Vienna, Vienna 1180, Austria 
\and Research Institute of Physics, Southern Federal University, Rostov-on-Don 344090, Russia 
}

\authorrunning{Vorobyov}

\date{Received --- ; Accepted: 21.03.2016}

\abstract
{}
{We investigate the dynamics of gaseous clumps formed via gravitational fragmentation 
in young protostellar disks, focusing on the fragments that are ejected from the 
disk via many-body gravitational interaction. }
{Numerical hydrodynamics simulations were employed to study the evolution of young protostellar disks
formed from the collapse of rotating pre-stellar cores with mass in the 1.1-1.6~$M_\odot$ range.}
{Protostellar disks formed in our models undergo gravitational fragmentation driven by
continuing mass loading from parental collapsing cores. A few fragments can be ejected from the disk during the early evolution, but the low-mass fragments ($< 15~M_{\rm Jup}$) 
disperse creating spectacular bow-type structures while passing through the disk and collapsing core.
The least massive fragment that survived the ejection ($21~M_{\rm Jup}$) straddles the planetary-mass limit,
while the most massive ejected fragments ($145~M_{\rm Jup}$) can break up into several pieces, leading
to the ejection of wide separation binary clumps in the brown-dwarf mass range. About half of the ejected fragments are
gravitationally bound, the majority is supported by rotation against gravity, and 
all fragments have the specific angular momentum  that is
much higher than that expected for brown dwarfs. We found that the internal structure of the
ejected fragments is  distinct from what would be expected for gravitationally contracting clumps formed via cloud core fragmentation,
which can help to differentiate their origin.}
{The ejection of fragments is an important process inherent
to  massive protostellar disks,
which produces freely-floating pre-brown dwarf cores, regulates the disk and stellar masses, and
potentially enriches the intracluster medium with processed dust and complex organics.}


\maketitle

\section{Introduction}
\label{intro}
Protostellar disks are formed during the gravitational collapse of rotating cloud cores
owing to conservation of angular momentum. During the early evolution phase when the disk 
is still embedded in the parental core, 
the mass infall rate onto the disk from the core often exceeds the disk 
mass loss onto the star via viscous or magnetic 
torques, leading to the net growth of the disk mass and development of gravitational instability
\citep[e.g.][]{VB2010, Machida2010}.  If the disk mass becomes sufficiently high 
($\ga 0.1~M_\odot$) and radiative cooling is sufficiently fast, the latter is usually the case 
at radii greater than a few tens of AU, then compact gaseous clumps can
form in densest regions of the disk via gravitational fragmentation 
\citep[e.g.][]{Gammie2003,Rafikov2007,Boley2009,VB2010,Meru2012,Tsukamoto2013}.
These fragments are supported against self-gravity by pressure forces and/or rotation and
have a mass spectrum ranging from a few Jupiter masses to upper brown dwarfs and very-low-mass 
stars \citep[e.g.][]{VZD2013,Galvagni2014}.  

The subsequent evolution of the clumps depends on a number of factors, but can be roughly divided  
into three main pathways. Many clumps that form in the embedded phase of star formation 
experience fast inward migration and may be tidally dispersed and accreted into the star 
producing luminosity bursts similar in magnitude to those 
of FU-Orionis-type objects \citep{VB2010,VB2015,Machida2011}, or may deliver a population 
of giant planets (including hot Jupiters), icy giants, 
or terrestrial planets to the inner disk if the clump contraction rate and dust sedimentation are 
sufficiently fast to withstand the tidal torques
\citep{Nayakshin2010,Boley2010,Galvagni2014}. 
Clumps that form in the late embedded phase or even in the early T Tauri phase usually 
experience slow inward or even outward migration and can form a  population of giant planets and brown
dwarfs on wide orbits \citep{VB2010,Zhu2012,Vorobyov2013,Stamatellos2015}.

Some clumps may be tidally destroyed during close encounters with other clumps or dense spiral 
arcs  \citep{Zhu2012} and, depending on the rate of dust 
sedimentation, may either release processed dust, such as 
crystalline silicates, or solid protoplanetary cores at various radii in the disk 
\citep{Boley2010,Vorobyov2011, Nayakshin2011}. Finally, if several clumps are present in the disk 
at a time, one of them can  be ejected into the intracluster medium via many-body gravitational 
interaction, producing a population of freely-floating objects which slowly cool and contract to
form brown dwarfs and very-low-mass stars \citep{Thies2010,BV2012}.

In this paper, we focus on the ejection of clumps from protostellar disks. 
We revisit the earlier work of \citet{BV2012} using numerical hydrodynamics simulations 
with higher numerical resolution, which allows us to better resolve the ejected clumps,
calculate their internal structure and consider possible implications of the ejection 
process for the evolution of disks and host stars. The properties of ejected clumps can help to
clarify the origin of proto-brown dwarf candidates recently detected 
in young star-forming regions \citep{Andre2012,PhanBao2014,Palau2014,Riaz2015}
The paper is organized as follows. The numerical model is briefly described in Section~\ref{model},
disk fragmentation and ejection of clumps are considered in Section~\ref{eject}, properties
of the clumps are presented in Section~\ref{clump_prop}, implications of clump ejection
are discussed in Section~\ref{implications}, and main results are summarised 
in Section~\ref{summary}.

\section{Model description}
\label{model}
We compute the dynamics of gravitationally unstable disks using the
numerical model described in detail in \citet{VB2010}. Here, we provide only a brief description 
for the reader's convenience. Numerical simulations start from a collapsing pre-stellar core 
of a certain mass, angular momentum and temperature, continue into the disk formation phase, and finish
in the T Tauri phase. The use of the thin-disk
limit allows us to follow the evolution of circumstellar disks on time scales longer than
allowed in full 3D simulations.

We solve the mass, momentum, and energy transport equations using vertically integrated
quantities. Disk self-gravity is calculated by solving for the Poisson equation 
using a two-dimensional Fourier convolution theorem for polar coordinates. The energy equation
includes compressional heating, heating due to stellar and background irradiation, radiative
cooling from the disk surfaces, and heating due to turbulent viscosity (parameterized 
via the usual Shakura \& Sunyaev prescription with $\alpha=0.005$).
A diffusion approximation to link the effective surface temperature of the disk 
with the midplane temperature is adopted and a smooth transition is 
introduced between the optically thin and optically thick regimes.
The stellar irradiation is based on a luminosity that is a
combination of accretion luminosity and photospheric luminosity, the latter
calculated from the pre-main-sequence tracks of \citet{DAM94} using the
current stellar mass. Frequency-integrated opacities are adopted from the 
calculations of \citet{BL94}. The vertically integrated pressure is related to the 
internal energy per surface area through an ideal gas
equation of state.

Models presented in this paper are run on a polar coordinate ($r,\phi$)
grid with $1024 \times 1024$ zones. The radial grid zones 
are logarithmically spaced, while the grid spacing in the azimuthal direction
is equidistant. A central sink cell of radius 20 AU
is employed\footnote{In this work, a larger than usual sink cell is used to increase
the spatial resolution on the logarithmically spaced grid in the outer disk regions. } 
and the potential of a central point mass is added to the 
disk self-gravity once a central star is formed. 
 
The innermost cell outside the central sink has a size of 0.13--0.14~AU depending 
on the radius of the computational region (i.e., on the cloud core radius). 
The latter varies in the 0.07--0.11~pc (14000--22000~AU) 
limits. The radial and azimuthal resolution are 0.7~AU at a distance of 100 AU
and 3~AU at a distance of 500~AU from the star, which is about a factor of 2 higher
than in the study of \citet{BV2012}. 

The Truelove criterion states that the local Jeans length must
be resolved by at least four numerical cells to correctly capture
disk fragmentation (Truelove et al. 1998).
The Jeans length in a self-gravitating thin disk can be calculated as \citep{Vorobyov2013}
\begin{equation}
R_{\rm J}={\langle v^2 \rangle \over \pi G \Sigma}, 
\end{equation}
where $\langle v^2 \rangle$ is the gas velocity dispersion,
$G$ the gravitational constant, and $\Sigma$ the gas surface density 

\begin{table}
\begin{center}
\caption{Initial model parameters}
\label{table1}
\renewcommand{\arraystretch}{1.5}
\begin{tabular*}{\columnwidth}{ @{\extracolsep{\fill}} c c c c c c}
\hline \hline
model & $M_{\rm cl}$ & $\Sigma_0$ & $\Omega_0$ & $r_0$ & $\beta$  \\
\hspace{1cm} &  ($M_\odot$)  & (g~cm$^{-2}$) & (km~s$^{-1}$~pc$^{-1}$) & (AU) & (\%)     \\ [0.5ex]
\hline \\ [-2.0ex]
1 & 1.23 & $4.5\times10^{-2}$ & 1.5 & 2745 & 1.25 \\
2 & 1.38 & $4.0\times10^{-2}$ & 1.3 & 3085 & 1.25  \\
3 & 1.07 & $5.2\times10^{-2}$ & 2.3 & 2400 & 2.25  \\
4 & 1.69 & $3.3\times10^{-2}$ & 1.1 & 3770 & 1.25 \\
5 & 1.69 & $4.5\times10^{-2}$ & 1.4 & 2770 & 2.2  \\
\hline
\end{tabular*}
\end{center}
\medskip
\end{table} 

\begin{figure*}
 \centering
  \includegraphics[width=13cm]{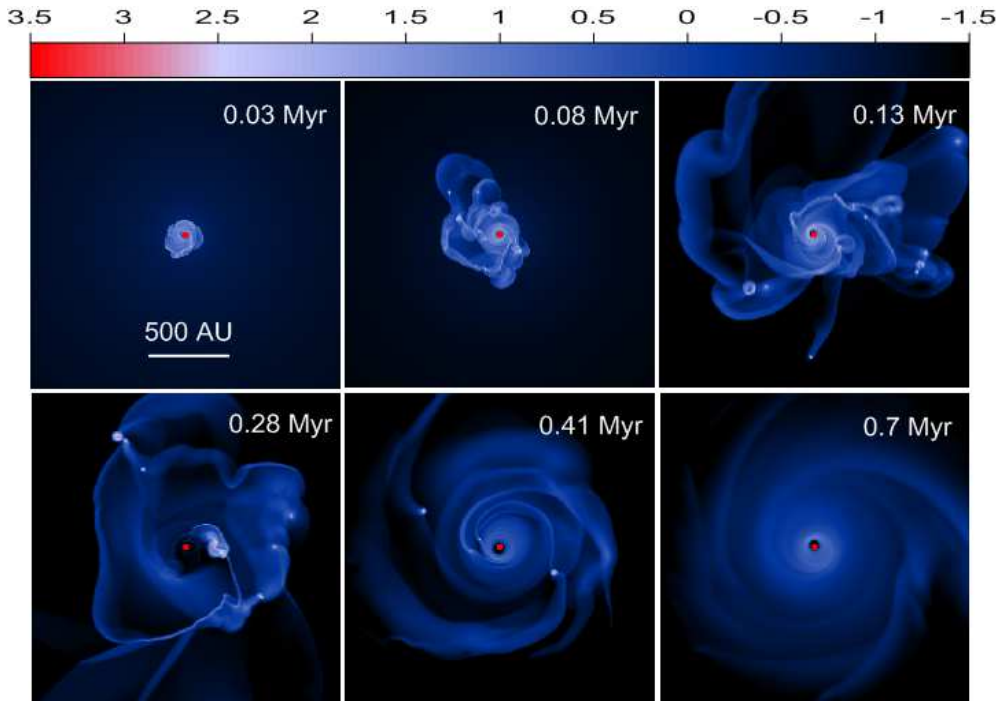}
  \caption{Evolution of the disk in model~1. Shown is the gas surface density in g~cm$^{-2}$ at six
  times elapsed since the formation of the central star (represented schematically by the red circle
  in the coordinate center). The scale bar is in log g~cm$^{-2}$. }
  \label{fig1}
\end{figure*}

Fragments usually form at distances greater than 50~AU. The 
typical surface densities and temperatures at 100~AU are $\Sigma=10$~g~cm$^{-2}$ and 
$T=25$~K \citep{Vorobyov2011}. Noting that $\langle v^2 \rangle=2 {\cal R} T/\mu$ for a thin disk
with two translational degrees of freedom,  where ${\cal R}$ is the universal gas constant and
$\mu=2.33$ the molecular wight, the corresponding Jeans length is about 50~AU. 
Since the gas surface density declines with radius faster than the gas temperature, $\Sigma~\propto r^{-1.5}$ and $T~\propto r^{-0.5}$ for self-gravitating flaring disks illuminated by stellar irradiation
\citep{Vorobyov2010}, the Jeans length is expected to increase with radius.
Our numerical resolution is therefore 
sufficient to resolve the Jeans length at distances $\la 1000$ AU, which is smaller than 
the maximum disk size in our numerical simulations.

The inner and outer boundary conditions are set to allow for free 
outflow from the computational domain.
The initial conditions for the pre-stellar core correspond to profiles
of column density $\Sigma$ and angular velocity $\Omega$ of the form
\barray
\Sigma & = & {r_0 \Sigma_0 \over \sqrt{r^2+r_0^2}}\:, \\
\Omega & = &2\Omega_0 \left( {r_0\over r}\right)^2 \left[\sqrt{1+\left({r\over r_0}\right)^2
} -1\right],
\label{ic}
\earray
where $\Sigma_0$ and $\Omega_0$ are the gas surface density and angular velocity 
at the center of the core. These profiles have a small near-uniform
central region of size $r_0$ and then transition to an $r^{-1}$ profile;
they are representative of a wide class of observations and theoretical models
\citep{Basu97,Andre93,Dapp09}. 
We have considered five representative models which show disk fragmentation and ejection
of fragments into the intracluster medium. The model parameters are
shown in Table~\ref{table1}, where $M_{\rm c}$ is the initial core mass and $\beta$ 
is the ratio of rotational energy to the magnitude of gravitational potential energy.
In all model cores, the initial gas temperature is set to 10~K.
The parameters of the models were chosen based on our previous experience 
to produce gravitationally unstable disks prone to fragmentation \citep{VB2010,BV2012}.

\section{Disk fragmentation and ejection of gaseous clumps}
\label{eject}
We start with analysing the long-term evolution of gravitationally unstable disks which
form from the gravitational collapse of our model cores.  Figure~1
presents the gas surface density distribution in model~1 at various times since the 
formation of the central star. We focus
on the inner computational region with a size of $2000 \times 2000$~AU$^2$ to better demonstrate
the complexity of the disk structure and dynamics during the early disk evolution. The disk 
forms at $t=0.01$~Myr (hereafter, the time is counted from  the formation of the central star) 
and quickly gains mass 
thanks to continuous mass loading from the infalling parental core.  The disk mass exceeds 
$0.1~M_\odot$ after $t=0.03$~Myr leading to the development of vigourous gravitational instability
and fragmentation. 
Gravitationally-bound clumps that form in the disk are essentially the ``first hydrostatic cores'' 
in the parlance of star formation \citep{Larson69} with typical sizes of several tens of AU and temperatures
up to several hundreds Kelvin \citep{VZD2013}. A second collapse
down to planetary-sized bodies can take place when the central temperature
reaches $\sim 2000$ K and molecular hydrogen begins to dissociate. This possibility is however not 
allowed in our numerical simulations, because resolving planetary-sized objects 
requires a local spatial 
resolution $\ll 0.1$~AU and is a formidable task for grid-based codes designed to study the 
{\it global} evolution of protostellar disks. We note however that in our simulations the majority
of fragments are destroyed or ejected from the disk before reaching 
the temperature needed for $H_2$ dissociation.

\begin{figure*}
 \centering
  \includegraphics[width=13cm]{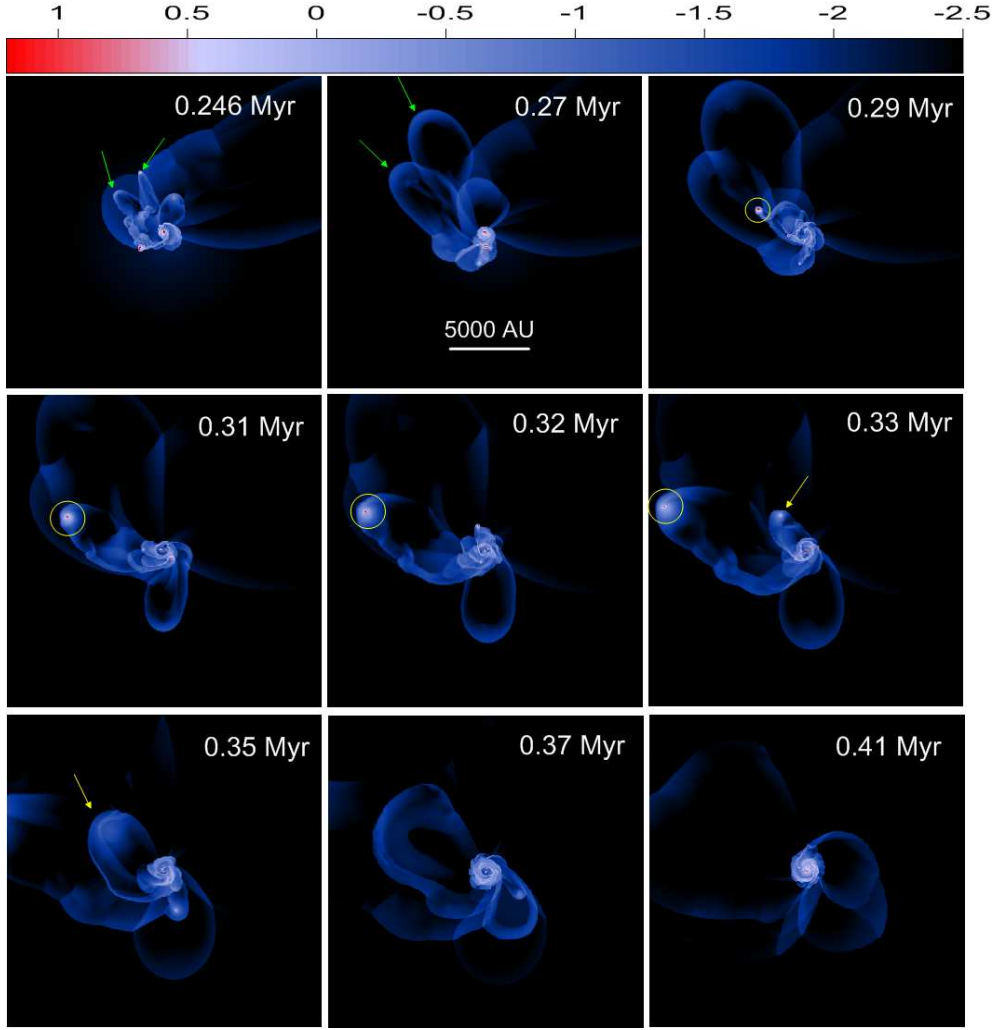}
  \caption{Three ejection events in model~1. The first ejected objects are shown with the 
  green arrows indicating low-mass clumps which dispersed soon after leaving the disk. 
  The yellow circles outline the second ejected object - a massive fragment
  that withstood ejection and left the computational region. The yellow arrows show the final 
  ejected clump which also dispersed. The vertical bar is in log g~cm$^{-2}$.}
  \label{fig2}
\end{figure*}

After $t=0.13$~Myr, the disk mass exceeds 0.2~$M_\odot$ and during the subsequent evolution the disk
reveals a very chaotic structure with multiple fragments connected by elements of spiral arms and arcs.
The fragments span a mass range from a few Jupiters masses to upper-mass brown dwarfs and
they are found at various distances in the disk, often embedded in spiral arms in relative isolation
from other fragments or in closely packed groups. After $t=0.41$~Myr, the disk
starts to take a more regular shape with a well-defined multi-arm spiral pattern. Several fragments
are still present in the disk at this late evolution time. However, 
after $t=0.7$~Myr all fragments have disappeared either migrating through the inner sink cell (onto probably onto the star) or being destroyed by tidal torques. These two fragment destruction
mechanisms have also been reported in previous independent numerical  studies 
\citep[e.g.][]{Cha11,Baruteau2011,Zhu2012,Tsukamoto2013}

In this work, we focus on yet another mechanism that can lead to the loss of fragments by 
protostellar disks -- ejection of fragment via multi-body interaction. The complexity of the 
disk structure during the early disk evolution (see Fig.~\ref{fig1}) implies the presence
of strong dynamical interactions between various disk components. If two (or more) fragments
experience a close approach with each other, this may lead to ejection of one of the fragments
into the intracluster medium via multi-body gravitational interaction. This effect was 
demonstrated in \citet{BV2012} in the context of self-gravitating fragments and earlier by, e.g., 
\citet{Stamatellos2009} in the context of fully formed brown dwarfs. 
The difference between these two cases is not trivial. In disk simulations, 
the fully formed brown dwarfs are usually described by point-sized
gravitating particles with a smoothing length of the gravitational potential on the order of 1~AU.
On the other hand, self-gravitating fragments in our study are at least an order of magnitude larger.
This makes it more difficult for the fragments to be ejected owing to a larger effective distance
between these objects during close encounters. Moreover, the finite-sizes fragments can
be destroyed via tidal torques during close encounters, which is not possible for point-sized
objects.

\begin{figure*}
 \centering
  \includegraphics[width=14cm]{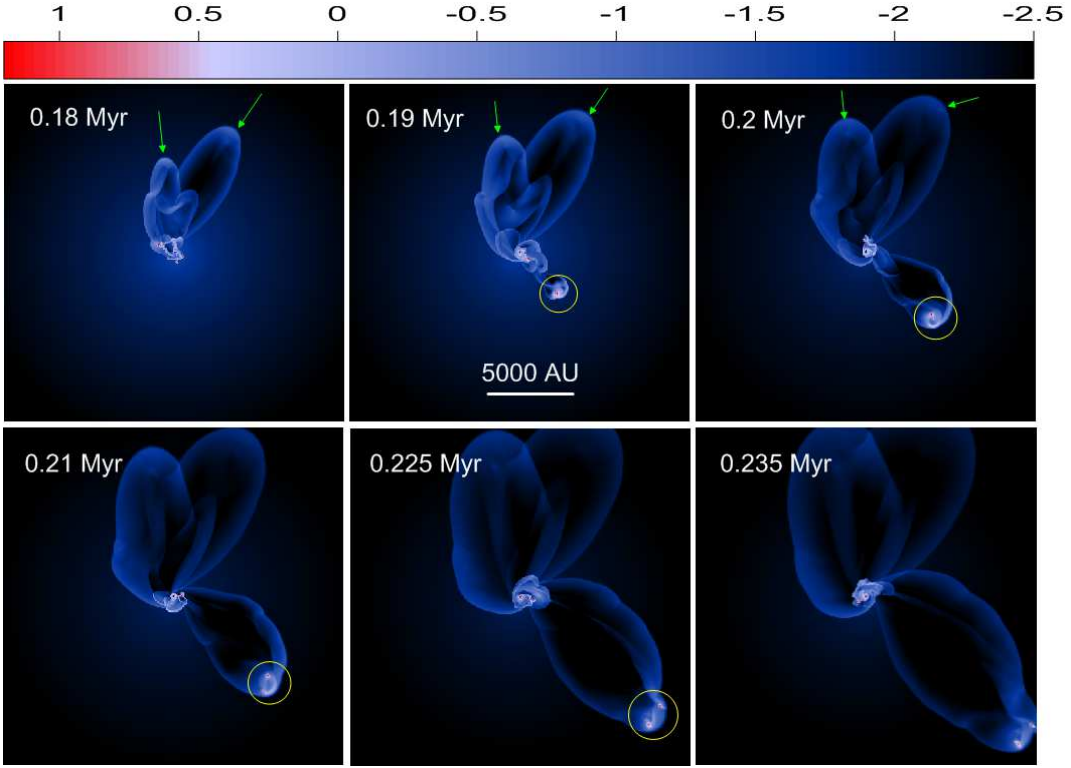}
  \caption{Two ejection events in model~2. The first ejected objects are shown with the 
  green arrows indicating low-mass clumps which dispersed soon after leaving the disk. 
  The second ejected object is a massive fragment outlined by the yellow circles,
   it breaks into a binary system soon after leaving the disk. The scale bar is in log g~cm$^{-2}$.}
  \label{fig5}
\end{figure*}

Figure~\ref{fig2} presents a zoomed-out view on the computational region in model~1, this time covering
a ten times larger box $20000\times20000$~AU$^2$ than in Figure~\ref{fig1}, but focusing 
on a shorter evolution sequence between $t=0.246$~Myr
and $t=0.41$~Myr.  We note that the entire computational region is still somewhat larger,
$30000\times30000$~AU$^2$, but we focus on a smaller region to allow for a better 
resolution of the inner regions.  The figure captures three ejection events described below in more
detail.

The first event occurs at $t\approx0.25$~Myr and 
is associated with ejection of several diffuse gaseous clumps indicated in the image with 
the green arrows. The masses of the clumps are below $10~M_{\rm Jup}$.
As these clumps traverse the disk with velocity $\approx1.0$~km~s$^{-1}$, more than a 
factor of 2 higher than the local speed of sound $\approx 0.4$~km~s$^{-1}$, 
they create spectacular bow shocks. 
The ram pressure from the disk and envelope ultimately destroys the clumps, but the 
material entrailed by the passage of the clump continues to expand and leaves the shown area 
at around $t=0.3$~Myr.

The second event is the ejection of a massive fragment ($75~M_{\rm Jup}$) outlined in 
Figure~\ref{fig2} by the yellow circles. The event takes place at around $t=0.28$~Myr. 
The ejected fragment is significantly more massive than the two clumps ejected during the first event.
The increased self-gravity of the massive fragment makes it easier to withstand the 
the destructive effect of tidal torques during the close encounter.
Moreover, the massive fragment happened to be ejected in almost the same direction 
as the previously ejected 
low-mass clumps, which effectively cleared the way for the ejected fragment so that the 
ram pressure of the envelope material became weaker.
The fragment passes through the boundary at around $t=0.33$~Myr having the velocity 
a factor of 1.8 higher than the escape velocity ($v_{\rm esc}=0.4$~km~s$^{-1}$).
This indicates the true ejection into the intracluster medium 
with little chance for the fragment to return.

Finally, the third ejection event takes place at $t\approx0.33$~Myr, but the ejected clump 
shown in Figure~\ref{fig2} by the yellow arrows does
not survive for long and disperses soon after leaving the disk. The entrailed material however
continues to expand and at least a fraction of it passes through the outer computational boundary.

\begin{figure}  
  \resizebox{\hsize}{!}{\includegraphics{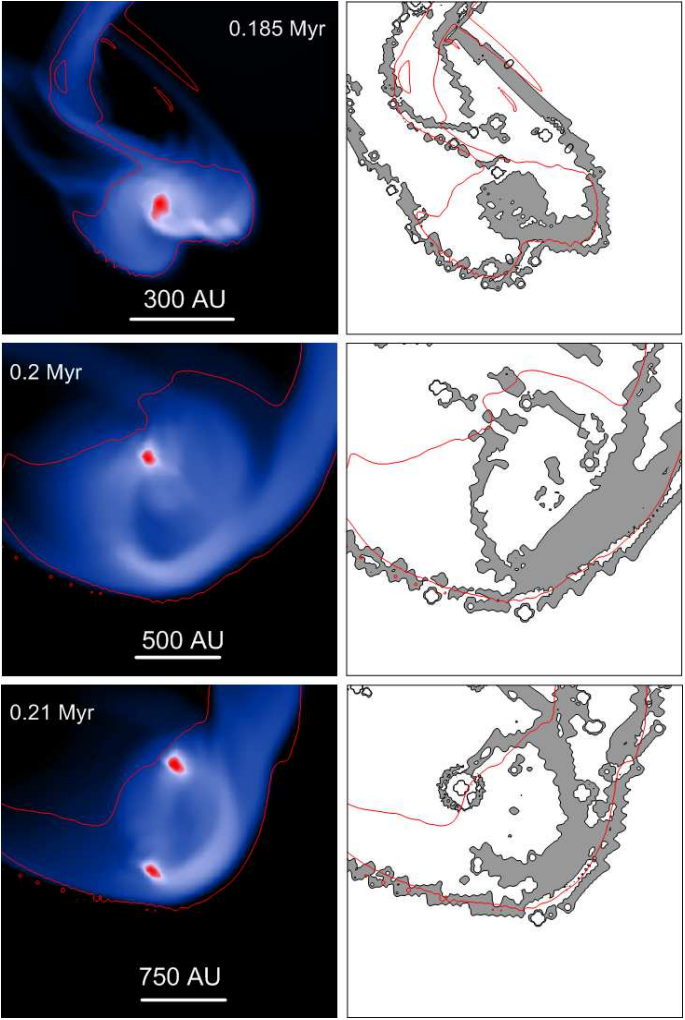}}   
    \caption{{\bf Left column} Distribution of the gas surface density (left column, g cm$^{-2}$, 
    log units) and the $\cal G$-parameter (right column, log units) in and around the ejected
fragment undergoing fragmentation.
Only the regions with ${\cal G}<1.0$ are shown with the grey color to identify areas where cooling 
is sufficiently fast for gravitational fragmentation to occur.
The contour lines outline the regions where the $Q$-parameter is  lower than unity.} 
    \label{fig6} 
 \end{figure}

\subsection{Ejection of a binary fragment}
The ejection events that occur in massive circumstellar disks are not limited to 
ejection of single fragments. The ejected fragments often possess sufficient mass and 
angular momentum to undergo internal fragmentation. This phenomenon 
was first reported in \citet{Vorobyov2013b}, but one of the fragments constituting
the ejected binary system dispersed
before leaving the computational region. A better numerical resolution 
of the current numerical simulations allows us to study this process in more detail.

Figure~\ref{fig5} presents a sequence of time snapshots in model~2 illustrating the ejection events
taking place around $t=0.18$~Myr. Each panel shows the gas surface density in the inner 
$20000\times20000$~AU$^2$ box. Similar to model~1, the green arrows track diffuse
clumps that disperse after being ejected from the disk, suggesting that this phenomenon 
is typical for massive disks. The yellow circles, on the other hand, show the ejected clump which survives
the ejection process but breaks into two separate fragments on its journey outward.
The original mass of the ejected clump was $\approx80~M_{\rm Jup}$ and the final masses of the
binary components are $33~M_{\rm Jup}$ and $30~M_{\rm Jup}$.

Two criteria are usually invoked to check the feasibility of gravitational fragmentation:
\begin{equation}
Q_{\rm T}={c_{\rm s} \kappa \over {\pi G \Sigma}} \le Q_{\rm crit} \,\,\,\, \mathrm{and} \,\,\,\,
{\cal G}=\Omega t_{\rm c} \le \cal{G}_{\rm crit},
\end{equation}
where $Q_{\rm T}$ is the Toomre parameter \citep{Toomre64}, $\kappa$ the epicyclic frequency, and 
$\cal G$ the product  of the local cooling time
$t_{\rm c}$ and orbital frequency $\Omega$ \citep{Gammie2001}. 
Gravitational fragmentation is supposed to occur if both parameters, $Q_{\rm T}$ and ${\cal G}$ 
are smaller than some critical values, $Q_{\rm crit}$ and ${\cal G}_{\rm crit}$.
There is an ongoing debate as to the exact values of $Q_{\rm crit}$ and ${\cal G}_{\rm crit}$
\citep[see e.g.][]{Meru2012,Rice2014}, but most studies seem to indicate 
that disk fragmentation occurs when both values are smaller than 1.0.

To check if the two fragmentation criteria are fulfilled in model~2,
we plot in Figure~\ref{fig6} the spatial distribution of the $Q$- and $\cal G$-parameters 
around the ejected clump. More specifically, the color image in the left column
presents the gas surface density distribution,
while the grey image in the right column shows the regions with ${\cal G}<1.0$
In both columns, the red lines outline the regions with $Q<1.0$.
Evidently, the $Q$-parameter is smaller than unity almost everywhere within the clump.
We note that when calculating the value of $Q$ we calculated the epicyclic frequency $\kappa$ 
in the frame of reference of the central gravitating star, while it would be 
more consistent to find the value of $\kappa$ in the rotating frame of reference of the ejected clump,
at least in the pre-fragmentation phase. However, the gas flow in and around the ejected 
fragment was
too complicated to derive reliable values of $\kappa$ in the rotating frame of reference, 
forcing us to use a simpler calculation of $\kappa$ in the fixed frame of reference.
The situation with the ${\cal G}$-parameter is more complicated, but there are areas
in the clump where both criteria for fragmentation are fulfilled, e.g., in
a spiral arc extending to the bottom-right, within which the second fragment finally
forms.

\subsection{The ejected fragment straddling the planetary-mass limit}
\begin{figure*} 
\centering 
  \includegraphics[width=14cm]{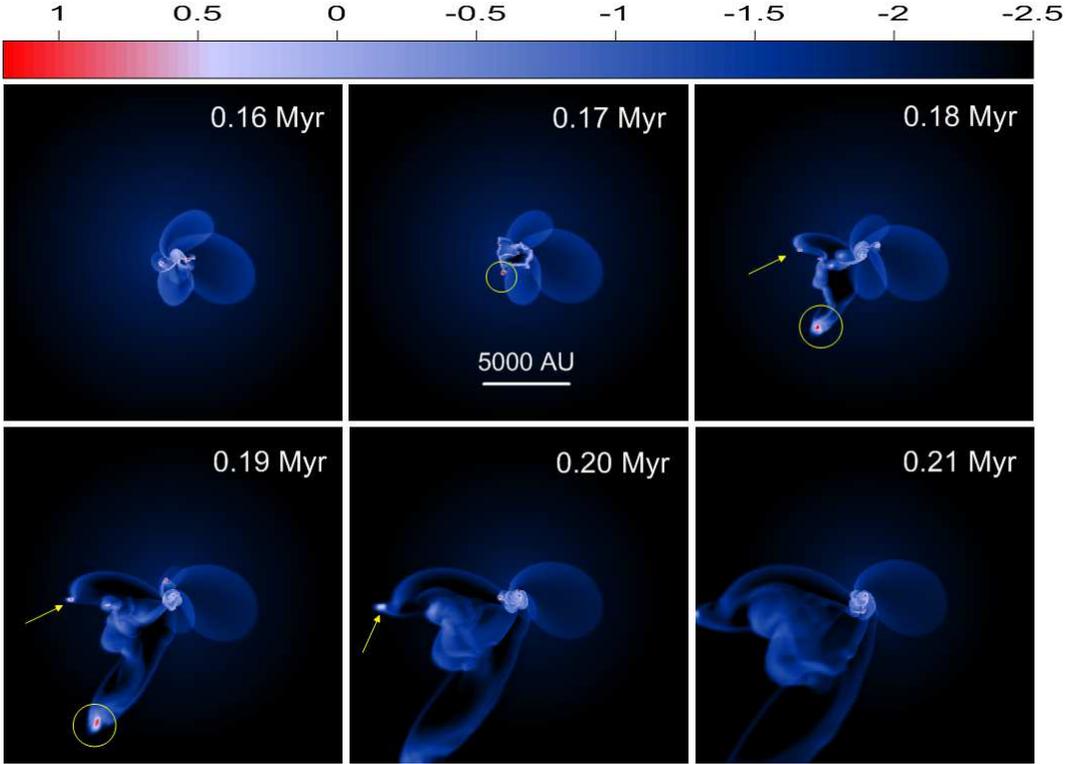}
    \caption{Two ejected events in model~3. The first ejected objects  are shown by the yellow and green
    arrows. They survive the ejection event and the leave the computational region. In particular,
    the fragment shown by the yellow arrow has the lowest mass found in our simulations, 
    21~$M_{\rm Jup}$. The second ejection is highlighted by the yellow circle comprising three fragments.     None of these objects survived the ejection. The scale bar is in g~cm$^{2}$. } 
    \label{fig7} 
 \end{figure*}

In the previously considered models, the ejected fragments had masses spanning a range 
from intermediate- to upper-mass brown dwarfs. In this section, 
we present the ejection of a fragment with mass straddling
the planetary-mass limit. Figure~\ref{fig7} shows the disk evolution in model~3 during a vigorous
ejection event taking place at $t\approx0.18$~Myr. Two ejected clumps identified
by the yellow and green arrows survive through the entire journey to the outer boundary
of the computational domain. Their masses
are $157~M_{\rm Jup}$ (green arrows) and $21~M_{\rm Jup}$ (yellow arrows).
Among all objects that survived the ejection process in our numerical simulations, 
the latter fragment is characterized by the lowest mass. 

The yellow circle in Figure~\ref{fig7} outlines another three ejected clumps with
masses ranging from 5 to 15~$M_{\rm Jup}$. They did not survive. 
It is possible that a weaker self-gravity
of low-mass clumps results in a reduced ability to withstand tidal torques acting on the clumps
during close encounters, leading to their subsequent dispersion. On the other hand,
similar numerical simulations of fragmenting disks with a twice
coarser numerical resolution \citep{BV2012} reported the minimum mass of ejected fragments to be 
about $40~M_{\rm Jup}$, which is almost a factor of 2 higher. 
This implies that the dispersal of low-mass fragments may be caused by the adopted log-spaced 
numerical grid, resulting in a progressively
coarser numerical resolution as the clumps propagate outward.
By extrapolation, we hope to detect
the ejection of planetary-mass objects in future numerical simulations with 
an increased numerical resolution.

\section{Properties of ejected fragments}
\label{clump_prop}

\begin{figure}
 \centering
  \resizebox{\hsize}{!}{\includegraphics{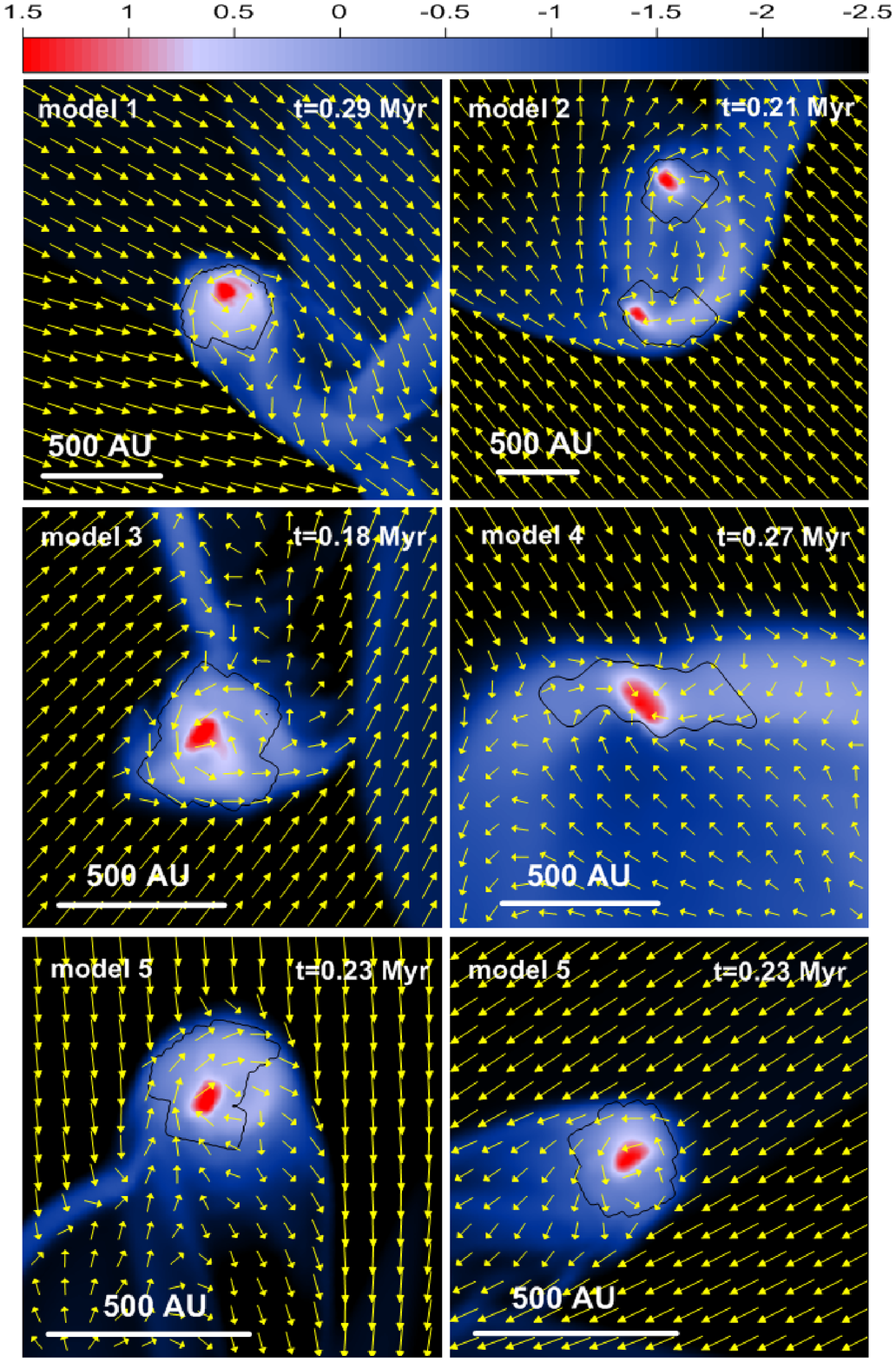}}
  \caption{Zoomed-in view on several ejected fragments in models~1-5. The corresponding time is indicated
  in every panel. Shown are the gas velocity field (yellow arrows) superimposed on 
  the gas surface density distribution. The black contour lines outline the fragments
  according to the adopted fragment tracing algorithm. The scale bar is  log g~cm$^{-2}$. }
  \label{fig4}
\end{figure}

The  numerical resolution of our models allow us to study in some detail the structure of ejected
fragments. Figure~\ref{fig4} presents a zoom-in view onto several ejected fragments in models~1-5.
In particular, the gas velocity field shown with the yellow arrows is 
superimposed on the gas surface density map in and around the ejected fragments. 
The velocity field is calculated in the framework of the moving fragment.  
The black lines outline the position of the fragments
according to the fragment tracking algorithm described 
in detail in \citet{Vorobyov2013}. This algorithm 
requires that the fragment be pressure-supported, having a negative
pressure gradient with respect to the center of the fragment, and that the fragment 
be kept together by gravity, having the deepest potential well at the center of the
fragment. These requirements translate into the following to equations:
\begin{eqnarray}
\label{pres}
{\partial {\cal P} \over \partial r^\prime} &+& {1 \over r^\prime}{\partial {\cal P} \over \partial \phi^\prime} <0, \\
{\partial \Phi \over \partial r^\prime} &+& {1 \over r^\prime}{\partial \Phi \over \partial \phi^\prime} >0,
\label{grav}
\end{eqnarray}
where  ($r,\phi$) are
the radial and azimuthal  coordinates of the polar grid, ($r_{\rm c},\phi_{\rm c})$ the radial 
and angular coordinates of the center of the fragment (defined as 
the local maximum in gas density), $r^\prime=r-r_{\rm c}$, $\phi^\prime=\phi-\phi_{\rm c}$, and $\Phi$ is the total gravitational potential including the 
contributions from the star, disk and envelope (if the latter is still present).

A visual inspection of Figure~\ref{fig4} indicates that the ejected fragments have a dense 
central core and a diffuse envelope.
The ejected fragments are also surrounded by a low-density material, which  is located inside
their Hill radii (greater than 1000 AU for most fragments, see Table~\ref{table2}) 
and is hence gravitationally bound to them. Several fragments have a 
cometary-like shape with  a compact head and a diffuse tail.
The ejected fragments may rotate not only in the same direction as the parental disk
(counterclockwise), but also in the opposite direction.

The main properties of the ejected fragments are listed in Table~\ref{table2}. Evidently,
the mass of most ejected fragments lies in the brown-dwarf limits. However, when the material within
the Hill radius is taken into account, the final mass for five objects shifts closer to very-low-mass
stars, ~0.1~$M_\odot$. On the other hand, because of high angular momentum these objects 
are expected to form disks and lose some
material via associated  jets and outflows, which may bring them back to the brown dwarf
mass range. We also note that all ejected objects have velocities greater than the escape speed, implying
that we witness here true ejections with little chance for the fragments to come back 
to the parental disk.

\begin{table*}
\center
\caption{Models showing ejection and properties of ejected clumps}
\label{table2}
\begin{tabular}{ccccccccccc}
\hline\hline
Model & time &  $M_{\rm fr}$ &  $M_{\rm Hill}$ & $R_{\rm Hill}$  & $v_{\rm fr}$ & $v_{\rm esc}$ & $E_{\rm rot}$ & $E_{\rm th}$
 & $E_{\rm grav}$ & $J_{\rm fr}$ \\
 & (Myr) & ($M_{\rm Jup}$) & ($M_{\rm Jup}$) & (AU)  & (km~s$^{-1}$) & (km~s$^{-1}$) & (g~cm$^2$~s$^{-2}$) & (g~cm$^2$~s$^{-2}$) & (g~cm$^2$~s$^{-2}$) & (cm$^2$ s$^{-1}$) \\
\hline

1& 0.29 & 75 & 84 & 1100 & 0.65 & 0.4 & $2.8\times10^{41}$ & $1.7\times10^{41}$ & -$7.4\times10^{41}$ & $4.5\times10^{19}$  \\
2a & 0.21  & 33 & 118 & 2000 & 0.95 & 0.43 & $1.8\times10^{40}$  & $5.7\times10^{40}$ & -$9.4\times10^{40}$ & $2.4\times10^{19}$  \\
2b & 0.21 & 30 & 120 & 1800  & 0.95 & 0.43 & $3.95\times10^{40}$  & $4.9\times10^{40}$ & -$1.2\times10^{41}$ & $3.3\times10^{19}$  \\
3a & 0.18 &  21 & 30 & 950  & 1.05 & 0.35 & $1.1\times10^{39}$ & $3.3\times10^{40}$ & -$5.4\times10^{40}$ & $5.5\times10^{18}$   \\
3b & 0.18 & 145 & 160  & 2330 & 1.8 & 0.35  & $1.0\times10^{42}$ &    $2.8\times10^{41}$ & -$1.6\times10^{42}$ & $1.6\times10^{20}$ \\
4a & 0.27  & 24 & 35 & 2020 & 1.3 & 0.4 & $1.4\times10^{39}$ & $3.7\times10^{40}$  & -$5.3\times10^{40}$ &  $8.9\times10^{18}$ \\
4b & 0.27& 59 & 151 & 1900 & 0.47 & 0.4  & $2.5\times10^{41}$ & $1.0\times10^{40}$  & -$3.5\times10^{41}$ & $1.2\times10^{20}$ \\
5a & 0.24 &  48 & 56 & 940  & 1.5 & 0.42  & $2.3\times10^{41}$ & $1.1\times10^{41}$ & -$4.9\times10^{41}$ & $4.1\times10^{19}$  \\
5b & 0.24 & 64 & 78 & 950  & 1.5 & 0.42 & $4.3\times10^{41}$ & $2.2\times10^{41}$ & -$8.2\times10^{41}$ & $5.5\times10^{19}$  \\
\hline
\end{tabular}\par
\vspace{5 pt}
{ $M_{\rm fr}$ is the mass of the fragment, $M_{\rm Hill}$ the mass within the Hill radius,
$R_{\rm Hill}$ the Hill radius, $v_{\rm fr}$ the fragment velocity, $v_{\rm esc}$ the escape velocity,
 $E_{\rm rot}$, $E_{\rm th}$, and $E_{\rm grav}$ the rotational, thermal, and gravitational 
 energies of the fragments, respectively, $J_{\rm fr}$ the specific angular momentum of the fragments.
 The second column gives the time when the fragment properties are calculated.}
\end{table*}

To better understand the evolution of the fragments,
we calculate their gravitational energy using the following equation:
\begin{equation}
E_{\rm grav} = - \int \Sigma  (r^\prime \cdot \nabla \Phi) dS,
\end{equation}
where the scalar product $(r^\prime \cdot \nabla \Phi)$ is calculated in the local frame of reference
defined by  $r_{\rm c}$ and
the integration is performed over all grid cells occupied by 
the fragment. The rotational and thermal energy are calculated as:
\begin{eqnarray}
E_{\rm rot} &=& {1\over 2} \int \Sigma v^2_\phi dS \\
E_{\rm th} &=& {{\cal R} \over (\gamma-1) \mu} \int \Sigma T dS,
\end{eqnarray}
where $v_\phi$ is the gas velocity determined in the local frame of reference of the fragment center,
$T$ the gas temperature, $\gamma=7/5$ the ratio of specific heats,
and $\mu=2.33$ the mean molecular weight.
Here, the integration is also performed over all grid cells occupied by 
the fragment.

The fragments are found to evolve near, but not exactly at the virial equilibrium.  
Four out of nine fragments (1, 2a, 3a, 4a) are gravitationally bound with 
$|E_{\rm grav}| \ga 2 E_{\rm rot}$  
+ $E_{\rm th}$, but others  are characterized by 
$|E_{\rm grav}| \la 2 E_{\rm rot}$  + $E_{\rm th}$, implying that they may disperse 
if they do not lose part of their thermal or rotational support during the subsequent evolution.
Three fragments out of nine (2a, 3a, and 4a) are mostly thermally supported against self-gravity, 
while the others are mostly supported by rotation. In all fragments, the specific angular momentum
is much higher than that expected for brown dwarfs with the rotation period of 1~day, $J_{\rm BD}= 4\times
10^{15}$~cm$^2$~s$^{-1}$.
This means that ejected fragments have to get rid of most angular momentum when contracting
into brown-dwarf-sized objects, possibly via the formation of circum-brown dwarf disks, jets and outflows.

\begin{figure}
 \centering
  \resizebox{\hsize}{!}{\includegraphics{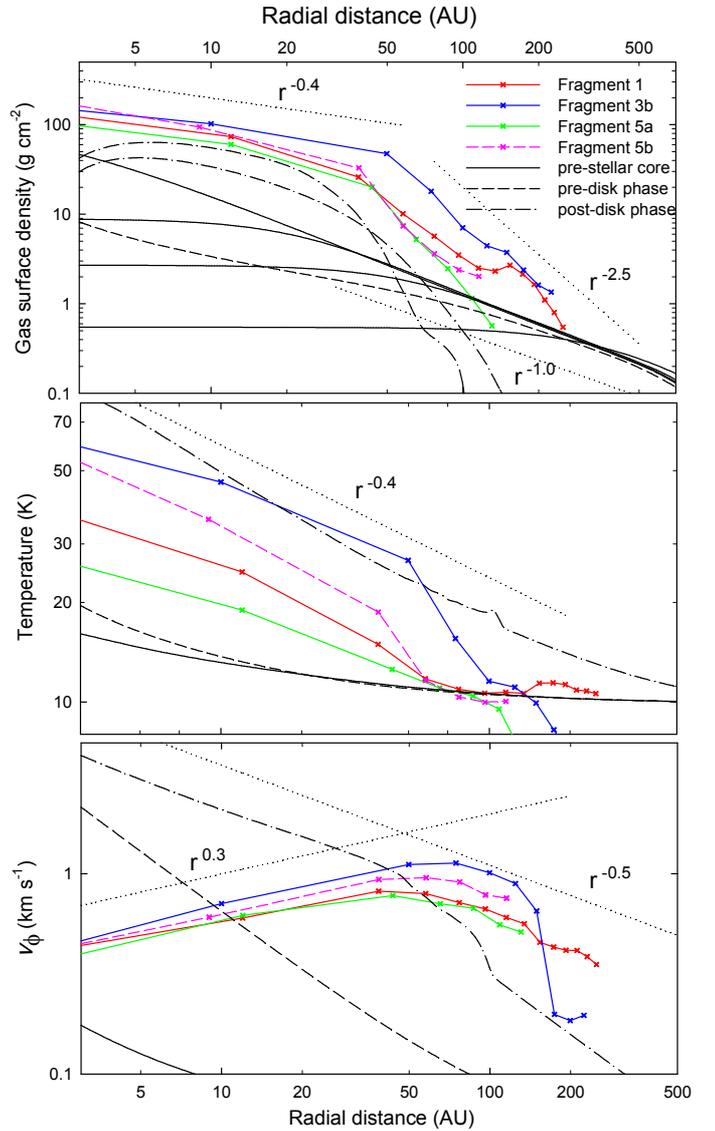}}
  \caption{Azimuthally averaged radial profiles of the gas surface
density (top panel), temperature (middle panel) and rotation velocity (bottom panel) 
for several best-resolved fragments in models~1, 3, and 5.  The black 
lines show the radial gas surface density, temperature and rotational
velocity  profiles of a vertically integrated, gravitationally contracting Bonnor-Ebert
sphere in the pre-stellar phase (solid lines), pre-disk phase (dashed lines), and post-disk phase
(dash-dotted lines). The doted lines present the radial slopes for comparison.}
  \label{fig9}
\end{figure}

Finally, in Figure~\ref{fig9} we present the azimuthally averaged radial profiles of the gas surface
density (top panel), temperature (middle panel) and rotation velocity (bottom panel) 
for several best-resolved fragments. Evidently, the fragments are characterized by a centrally 
condensed surface density distribution, which has a near-constant-density plateau 
($\propto r^{-0.4}$) with a size on the order of 50~AU and an extended tail that falls 
off with radius as r$^{-2.5}$. The radial temperature profiles decline with radius as $r^{-0.4}$,
but in general show a much wider scatter at a given radius than the surface density profiles. 
In contrast to the gas surface density and temperature distributions, the 
radial distribution of rotational velocity ($v_\phi$) does not show a gradual decline with radius, but 
first increases with radius as $r^{0.3}$, reaches a local peak at 40--70~AU, and then
falls off  with radius roughly as $r^{-0.5}$. The latter suggests that a larger fraction 
of the total fragment mass is concentrated in the central near-constant-density core.

It is interesting to  compare the gas surface density distribution in the ejected fragments 
with that of a vertically integrated Bonnor-Ebert (BE) sphere described by the following 
equation \citep{Dapp09}
\begin{equation}
\Sigma  =  {\Sigma_{\rm c} \over \sqrt{1+(r/r_{\rm c})^2}} \,\, \mathrm{arctan} 
\left( \sqrt{C^2 - (r/r_{\rm c})^2 \over 1 + (r/r_{\rm c})^2} \right),  
\end{equation}
where $\Sigma_{\rm c}= 2 r_{\rm c} \rho_0 $ is the central surface density, $r_{\rm c}= c_{\rm s}/
\sqrt{2\pi G\rho_0}$ the radius of the central constant density plateau, $\rho_0$ the central gas
volume density, $c_{\rm s}$ the sound speed and $C$ the so-called concentration parameter chosen to
be equal to 5. For $T=10$~K, $n_0=10^7$~cm$^{-3}$, and $\beta=0.95\%$, typical for dense pre-brown 
dwarf cloud cores, the total mass contained within 1000~AU is equal to 70~$M_{\rm Jup}$, 
which is comparable to the mass of the most massive ejected clumps. 
The corresponding surface density profiles 
are shown in Figure~\ref{fig9} by the solid black lines.   More specifically, the 
bottom line shows the initial radial distribution, while
the other black solid lines (from bottom to top) correspond to $t=0.01$~Myr,
$t=0.011$~Myr, and $t=0.0115$~Myr after the onset of gravitational collapse. 
We do not limit us with the initial distribution, but consider the dynamically evolving core,
because the radial profiles are expected to change with time \citep[e.g.][]{Masunaga1998}.
Clearly, the surface density distribution of
the vertically integrated, gravitationally contracting BE
sphere is quite distinct from that of the ejected clumps. The former has a 
significantly lower gas surface density between a few AU and few hundreds of AU than the latter. 
For a similar total gas mass, ejected clumps are more compact than collapsing BE
spheres. Moreover, the radial slope of the gas surface density is also distinct in 
the two cases: for the BE sphere it is $\Sigma\propto r^{-1}$  and for the ejected clumps it 
has a bi-modal distribution, $\Sigma \propto r^{-0.4}$ for the clump's core and 
$\Sigma ~\propto r^{-2.5}$ for the clump's envelope.

To make our comparison more comprehensive, we followed the evolution of our collapsing BE sphere 
into the pre-disk and post-disk formation phases. The dashed line presents the radial gas surface 
density distribution at $t=0.013$~Myr, when the mass in the central sink cell 
(with a radius of 2~AU for this model) has exceeded 0.015~$M_\odot$ and the density 
has exceeded the opacity limit of $10^{11}$~cm$^{-3}$ (implying 
the formation of the first hydrostatic core there). At this stage, the radial
surface density profile in the collapsing core is again very distinct 
from that of the ejected fragments. Only after the formation of a centrifugally balanced disk,
the resulting surface density profiles shown by the dash-dotted lines for $t=0.04$~Myr and 
$t=0.07$~Myr, start resembling those of the ejected fragments. At this stage, however, 
BDs should be relatively easy distinguished from the ejected fragments due to 
the presence of a compact  central object with mass $>0.05~M_\odot$.

The radial profiles of gas temperature and rotational velocity are also distinct in 
the two cases, as can be seen in the middle and bottom panels of Figure~\ref{fig9}.
This time, we plot the radial distributions of our model only at $t=0.0115$~Myr 
(final stage of the BE sphere collapse), at $t=0.013$~Myr (pre-disk phase), and at 
$t=0.04$~Myr (the early post-disk phase).  In general, the gas temperature in the 
gravitationally contracting BE sphere is lower than in 
ejected fragments, while it is higher in the post-disk phase. 
However, the largest difference between contracting BE spheres and
ejected fragments is found for the case of rotational velocities. The former have radially
declining rotational velocities, which are sub-Keplerian in the pre-disk phase 
and Keplerian in the disk phase. The latter are described by a bi-modal velocity distribution,
increasing with distance in the inner 40--70~AU and changing to a near-Keplerian profile 
at larger distances, which suggests
that the central regions of ejected fragments are supported against gravity mostly by gas pressure.

\section{Implications}
\label{implications}
The  process of ejection of gaseous clumps from gravitationally unstable
protostellar disks has several interesting implications which we briefly 
discuss below.

{\it The origin of brown dwarfs.}  The formation of brown dwarfs
via the direct gravitational collapse of dense cores formed in molecular clouds 
via, e.g., turbulent compression has been advocated by many theoretical
and observational studies \citep[e.g.][]{Padoan2004, Hennebelle2008,Bate2009,Bate2012,Andre2012}. 
While certainly presenting 
a feasible gateway, this star-like scenario is not without problems because it requires high gas 
densities in the center of pre-brown dwarf cores and some mechanisms to truncate the cores 
(otherwise, the brown dwarf embryo would continue growing via accretion to stellar masses). 
An alternative to 
the star-like mechanism of brown dwarf formation is the gravitational fragmentation of protostellar
disks followed by ejection of either finished brown dwarfs \citep{Stamatellos2009} 
or gaseous clumps, which are expected later to cool and form freely-floating brown dwarfs 
\citep{BV2012,Vorobyov2013b}. 

The current study suggests that the radial density, temperature, and velocity distributions 
of ejected clumps are distinct from those of gravitationally contracting Bonnor-Ebert spheres, which
are usually invoked to describe dense cores formed via molecular cloud fragmentation. 
The former are more compact and centrally condensed, have higher temperatures and rotation
velocities than the Bonnor-Ebert-like cores. In particular, some of the ejected fragments 
are rotationally supported, which is difficult to expect from  the pressure supported Bonnor-Ebert
spheres. The typical ratios of the rotational to gravitational energies in pre-stellar cores
formed via molecular cloud fragmentation, $\beta=10^{-4}-0.07$ \citep{Caselli02}, also 
argue in favour of predominant pressure support.
This difference in the structure of pre-brown dwarf cores can be used to observationally 
distinguish between different formation mechanisms of brown dwarfs.

\begin{figure}
 \centering
  \resizebox{\hsize}{!}{\includegraphics{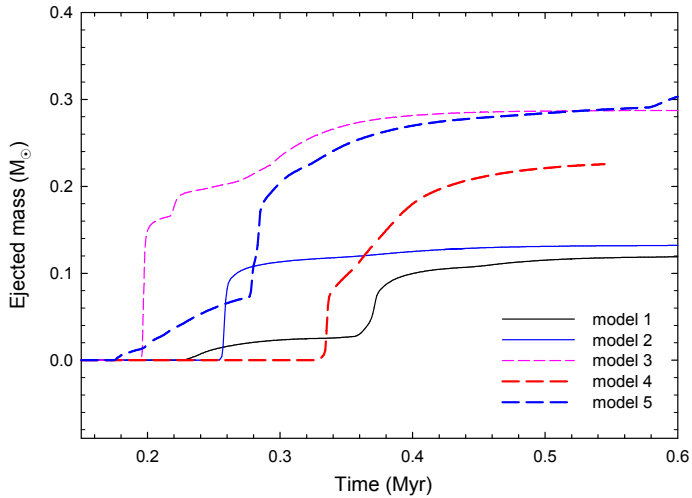}}
  \caption{Mass of gas that leaves the outer 
computational boundary in models~1-5 as a function of time elapsed 
since the formation of the disk. Steep rises in the outflown mass are caused
by dense fragments passing through the outer boundary after being ejected from the disk. }
  \label{fig10}
\end{figure}

{\it Stellar and disk masses}. 
The ejection of gaseous clumps from protostellar disks presents a substantial mass 
loss channel for young stellar systems, which can potentially affect the disk mass 
in the later evolution phase and the 
terminal mass of the star. Figure~\ref{fig10} shows the gas mass that leaves the outer 
computational boundary in five models as a function of time elapsed since the formation 
of the disk in each model. 
To remind the reader, we imposed a free outflow condition at the outer boundary, preventing 
any material to enter the computational domain so that the shown values represent the true ejected 
mass, not contaminated by any possible inflow. 

Compact fragments that survive through the ejection process cause a steep increase in the ejected mass
as they quickly pass through the outer computational boundary,
while diffuse clumps that disperse soon after leaving the disk create a gradual increase 
in the ejected mass because it takes a longer time for them to pass through 
the outer boundary. The total ejected mass ranges from $0.1~M_\odot$ to $0.3~M_\odot$, which
corresponds to 10\%--30\% of the total mass reservoir contained initially in parental cores 
or up to 50\% of the protostellar disk mass at the end of simulations. 
The ejection of gaseous clumps,
whether survived or dispersed, presents therefore an important mass regulation mechanism,
which limits the disk mass growth and reduces the terminal stellar mass. 
Although this channel alone is not sufficient to explain the observed shift between
the initial core and stellar mass functions \citep{Alves2007}, it may 
nevertheless present one of the possible processes that finally shapes
the initial mass function of stars.

{\it Enrichment of the intracluster medium}. Ejection of gaseous clumps  
presents an interesting mechanism for the enrichment of the intracluster medium
with the disk-processed material. Figure~\ref{fig10} shows that most ejection events 
take place between 0.2~Myr and 0.4~Myr. Protostellar disks with age of a few hundred thousand years
may have the dust content that is notably different from that of the interstellar medium 
\citep{Williams}. 
Moreover, gaseous clumps can act as dust attractors owing to the gas pressure gradients
\citep{Rice2004}, can anneal amorphous silicates into the crystalline form \citep{Vorobyov2011,NCB2011},  and can facilitate the growth of solid protoplanetary cores in their interiors 
\citep{Boss1998,Nayakshin2010,Boley2010,Nayakshin2011}.  If dispersed in the process of ejection, 
as was often seen in our numerical simulations, these clumps would release their dust content and
solid cores, enriching the intracluster medium with processed dust and creating a population of 
freely floating solid protoplanetary cores. Finally, the disk chemistry is also expected
to be different form that of the interstellar medium and the ejected clumps may 
enrich the intracluster medium with complex chemical elements (e.g., organics), which 
would be otherwise difficult to form. Further numerical studies are required to assess the 
importance and efficiency of the aforementioned effects.

\section{Conclusions}
\label{summary}

In this work, we have investigated numerically the long-term evolution 
of young self-gravitating protostellar disks exhibiting gravitational 
instability and formation of multiple gaseous fragments. We found that some of these 
fragments can be ejected
from the disk during close encounters with other fragments via the gravitational multi-body 
interaction. The least massive clumps ($\la 20 M_{\rm Jup}$) disperse while passing
through the disk and infalling envelope, creating spectacular bow-type structures.
More massive clumps, on the other hand, can survive the whole
journey to the outer computational boundary ($\ga 10000$~AU), producing a population of
freely-floating objects in the brown-dwarf mass range. 
Some of the most massive fragments can undergo gravitational fragmentation
after being ejected from the disk, leading to the ejection of binary objects with 
a separation of several hundreds of AU. All ejected clumps have velocities greater than the escape speed,
meaning that these are true ejections rather than mere scattering.

The minimum mass of the fragments that survived the ejection,  21~$M_{\rm Jup}$,
approaches the planetary-mass limit. This value is a factor of 1.5 smaller than what was 
found in a similar study by \citet{BV2012} employing a lower
numerical resolution. That suggests that the dispersal of clumps with mass $\la 20~M_{\rm Jup}$ 
is at least partly due to insufficient resolution on our logarithmically spaced polar grid.
We hope to see the ejection of planetary-mass objects in future higher-resolution simulations.

For about half of the ejected fragments, the sum of the thermal and rotational energies
is slightly greater than the gravitational energy, implying that they may disperse 
if they do not lose part of their thermal or rotational support during the subsequent evolution.
All clumps are characterized by the specific angular momentum that
is several orders of magnitude higher than expected for brown dwarfs, meaning that the ejected 
fragments have to get rid of most angular momentum when contracting into brown-dwarf-sized objects, possibly via the formation of circum-brown dwarf disks, jets and outflows.

We analyzed the internal structure of ejected fragments and found that it is
distinct from what would be expected for typical pre-brown dwarf cores 
formed via molecular cloud fragmentation.  
The ejected clumps are more centrally condensed and have a higher central temperature than
gravitationally contracting Bonnor-Ebert spheres.
The rotational velocity of the former has a bi-modal distribution and
is very distinct from the rotational pattern of the latter.  
This difference can be used for distinguishing the 
origin of brown dwarfs.  

We note that the ejected fragments can carry away up to 30\% of the total 
mass reservoir confined initially in parental cores, reducing the final stellar and disk 
masses and probably enriching the  intracluster medium with processed dust and complex
organics.
 
Finally, we note that in the current numerical simulations we have not taken into account
the finite contraction timescale of fragments. If the gas temperature exceeds 2000~K, 
dissociation of molecular hydrogen enables the contraction of AU-sized fragments 
to stellar-sized objects. This phenomenon is not captured  
in our numerical simulations due to the insufficient numerical 
resolution of our grid-based code, but may in fact reduce the efficiency of fragment ejection.
Numerical simulations employing sink particles as proxies for collapsed fragments
are planned to assess this effect.

\begin{acknowledgements}

The author is thankful to the anonymous referee for insightful comments and suggestions that
helped to improve the manuscript. Numerical simulations were done 
on the Atlantic Computational  Excellence Network (ACEnet), Shared Hierarchical
Academic Research Computing Network (SHARCNET), and Vienna Scientific Cluster (VSC-2).
This project was supported by the Russian Ministry
of Education and Science Grant 3.961.2014/K.  This publication is
supported by the Austrian Science Fund (FWF).

\end{acknowledgements}


\end{document}